\documentstyle[aps,epsfig]{revtex}

\begin{document} 
\title{The Type-problem on the Average for random walks on graphs}
\author{Raffaella Burioni\footnote{E-mail:
burioni@pr.infn.it}, Davide Cassi\footnote{E-mail: cassi@pr.infn.it},
Alessandro Vezzani\footnote{E-mail: vezzani@pr.infn.it}}
\address{Istituto Nazionale di Fisica della Materia, Dipartimento di
Fisica\\ 
Universit\`a di Parma, Parco Area delle Scienze n.7A, 43100 Parma, Italy  }
\date{\today}
\maketitle
\begin{abstract}
When averages over all starting points are considered, the
Type Problem for the recurrence or transience of a simple random walk on an 
inhomogeneous network in general differs from the usual "local" Type Problem.
This difference leads to a new classification of inhomogeneous discrete
structures in terms of {\it recurrence} and {\it transience} {\it on the average}, 
describing their large scale topology from a "statistical" point of view.
In this paper we analyze this classification and the properties connected to it,
showing how the average behavior affects the thermodynamic properties of 
statistical models on graphs. 
\end{abstract}
%%%%%%%

\section{Introduction}

In statistical mechanics and field theory Euclidean lattices describe the 
geometrical structure of crystals and of more abstract geometrical objects, 
such as 
discretized flat space-time. However, most real systems have irregular 
geometry: this  is the case for
glasses, polymers, amorphous materials, biological structures and fractals in condensed matter, as
well as discretized curved space-times in field theory.
The geometrical model for these inhomogeneous systems is  
a general discrete network made of sites and links, i.e.
a graph. From this point of view  usual lattices are a particular 
kind of graphs.
Statistical models defined on graphs (e.g. harmonic oscillations, random walks, 
spin system) are  the natural way to describe the physical 
properties of inhomogeneous real structures.

The study of the relation between  geometry and physics is one of the most 
complex and interesting problem of statistical mechanics and 
field theory on graphs. The main link between this two 
aspects is provided by random walks. 
The latter  are usually introduced to describe the diffusion of a classical 
particle and they  
are related to Markov chains, potential theory and algebraic graph theory 
on one side \cite{woess1}, and  to many problems of equilibrium 
and non equilibrium statistical mechanics, disordered systems and 
field theory on the other \cite{bg}.

In particular, the large times asymptotics of random walks provides the most 
effective method to describe the influence of large scale topology on the 
physical properties of discrete structures. The definition of the spectral
dimension for inhomogeneous networks, generalizing the Euclidean dimension
of lattices in field theory and phase transitions, is indeed based on long 
time behavior of random walks \cite{ao,monw,hhw}. More generally, this 
asymptotic regime allows to classify every graph either as {\it locally} 
{\it recursive} or 
{\it transient}, according to the probability
of ever returning to the starting site: the probability is 1 in 
the former case and less than 1 in the latter, independently of the site.
This classification, first introduced by Polya for regular lattices \cite{pol},
is known as the Type-Problem.

Local transience and recurrence describe
local properties of physical models on graphs.
However, in the study of statistical models on graphs we are in general 
interested in average (extensive) thermodynamic quantities. 
Indeed, while on lattices, due to translation invariance, local quantities 
are the same on all sites and therefore they are equal to their average, 
on inhomogeneous 
structures they depend  in general on the site and the average behavior 
can not be reduced to the local one.
In the last few years it has become clear that bulk properties are affected 
by the average values 
of random walks return probabilities over all starting sites: this is the 
case for spontaneous breaking of continuous symmetries \cite{mwg}, 
critical exponents of the spherical model \cite{cf}, harmonic vibrational 
spectra \cite{univ}. 
Therefore   
the classification of discrete structure in terms of {\it recurrence on 
the average} and {\it transience on the average} appears to be the most 
suitable. 
Unfortunately, while for regular lattices the two classifications are 
equivalent,
on more general networks they can be different and one has to study a 
Type-Problem
on the Average \cite{split}. 

Recently this problem has acquired particular relevance in the study of spin 
models on graphs. Indeed it has been shown that spontaneous breaking of
continuous symmetries occurs at $T>0$ if and only if the underlying
network is transient on the average \cite{bcv}. Moreover
this analysis has shown that relevant and new topological properties of 
infinite graphs are associated to the {\it on the average} classification.    

In this paper we deal with the Type-Problem on the Average and with the topological and 
thermodynamic properties arising from it, proving some basic theorems  
and discussing their relevance for present and future development 
in statistical physics. 

The paper is organized as follows. In the next section we introduce
the basic concepts and notations concerning random walks on infinite graphs.  
In section III we analyze local recurrence and transience properties 
of random walks, defined by the asymptotic behavior of return probabilities
generating function.
In section IV we consider thermodynamic averages on infinite graphs and in
section V  we introduce the topological classification in terms of 
average properties of random walks, based on transience and 
recurrence {\it on the average}.
In section VI we analyze the general relations between average 
generating functions of return probabilities and consider a further
classification holding for transient on the average graphs, which completes
the topological description of infinite graphs in terms of random walk
behavior. In section VII and VIII we show the relevance
of this topological classification in the study of thermodynamic
properties of statistical models on inhomogeneous structures.  
A summary and a discussion of our results are presented in section IX.

\section{Random walks on infinite graphs}

Let us begin by  recalling the basic definitions and results concerning graph theory and
random walks on infinite graphs, which will be used in the following.
A more detailed and complete treatment can be found in the mathematical 
reviews by Woess \cite{woess1,woess2}.
 
A graph $\cal{G}$ is a countable set $V$ of vertices (or sites)  $(i)$ 
connected pairwise by a set $E$ of unoriented links (or bonds) $(i,j)=(j,i)$.  
If the set $V$ is finite, ${\cal{G}}$ is 
called a finite graph and we will denote by $N$ the number of vertices of
$\cal{G}$. A subgraph $\cal{S}$ of $\cal{G}$ is a graph whose set of vertices
$S\subseteq V$ and whose set of links $E' \subseteq E$.

A path in $\cal {G}$ is a sequence of consecutive links 
$\{(i,k)(k,h)\dots(n,m)(m,j)\}$ 
and a graph is said to be connected, if for any two points $i,j \in V$ 
there is always a path 
joining them. In the following we will consider only connected graphs. 

The graph topology can be algebraically represented  
introducing its adjacency matrix $A_{ij}$ given by:
\begin{equation}
A_{ij}=\left\{
\begin{array}{cl}
1 & {\rm if } \ (i,j) \in E \cr
0 & {\rm if } \ (i,j) \not\in E \cr
\end{array}
\right .
\label{defA}
\end{equation}
The Laplacian matrix $\Delta_{ij}$ is defined by:
\begin{equation}
\Delta_{ij} = z_i \delta_{ij} - A_{ij}
\label{defDelta}
\end{equation}
where $z_i=\sum_j A_{ij}$, the number of nearest neighbors of $i$, 
is called the coordination number of site $i$. Here we will 
consider graphs with $z_{max}=\sup_{i} z_i < \infty$. 

In order to describe disordered structures we introduce 
a generalization of the adjacency matrix  given by the ferromagnetic
coupling matrix $J_{ij}$, with $J_{ij} \neq 0 \iff A_{ij}=1$ and
$\sup_{(i,j)} J_{ij} < \infty$, $\inf_{(i,j)} J_{ij} > 0$. 
One can then define the generalized 
Laplacian:
\begin{equation}
L_{ij} = I_i \delta_{ij} - J_{ij}
\label{defL}
\end{equation}
where $ I_i =\sum_j J_{ij}$.

Every connected graph $\cal{G}$ is endowed with an intrinsic metric 
generated by the chemical distance $r_{i,j}$  
which is defined as the number of links in the shortest path 
connecting vertices $i$ and $j$. 

Let us now introduce the random walk on a graph $\cal{G}$ defining the jumping 
probability $p_{ij}$ between nearest neighbors sites $i$ and $j$:
\begin{equation}
p_{ij}={A_{ij}\over z_i} =(Z^{-1}A)_{ij}
\label{pij}
\end{equation}
where $Z_{ij}= z_i \delta_{ij}$. From (\ref{pij}) the probability of 
reaching in $t$ steps site $j$ starting from $i$ is given by:
\begin{equation}
P_{ij}(t)=(p^t)_{ij}~~.
\label{ptij}
\end{equation}
Recurrence properties of random walks are studied introducing the 
probability $F_{ij}(t)$ for a walker starting from $i$ of reaching
for the first time in $t$ steps the site $j\not = i$, while $F_{ii}(t)$ 
is the probability of returning to the starting point $i$ for the first time 
after $t$ steps and $F_{ii}(0)=0$.  
The basic relationship between $P_{ij}(t)$ and $F_{ij}(t)$ is given by:
\begin{equation}
P_{ij}(t)=\sum_{k=0}^t F_{ij}(k) P_{jj}(t-k)
\label{PF}
\end{equation}
$(t>0)$. From the previous definitions $F_{ij} \equiv \sum_{t=0}^{\infty}
F_{ij}(t)$ turns out to be the probability of ever reaching the site $j$ starting from
$i$ (or of ever returning to $i$ if $j=i$). Therefore $0< F_{ij}\le 1$. 
The generating functions $\tilde{P}_{ij}(\lambda)$ and 
$\tilde{F}_{ij}(\lambda)$ are given by:
\begin{equation}
\tilde{P}_{ij}(\lambda)=\sum_{t=0}^{\infty}\lambda^t P_{ij}(t)
\ \ \ \tilde{F}_{ij}(\lambda)=\sum_{t=0}^{\infty}\lambda^t F_{ij}(t)~~
\label{genfun}
\end{equation}
where $\lambda $ is a complex number. From definition 
(\ref{genfun}) and from the property $0< F_{ij}\le 1$ by Abel theorem 
we have that $\tilde{F}_{ij}(\lambda)$ is a uniformly continuous function for 
$\lambda \in [0,1]$ and $0<\tilde{F}_{ij}(\lambda)\leq 1$, while 
$\tilde{P}_{ij}(\lambda)$ is continuous for $\lambda \in [0,1[$ but it can 
diverge for $\lambda\rightarrow 1^-$.

Multiply equations (\ref{PF}) by $\lambda^t$ and then summing over all 
possible $t$ with the initial condition $P_{ij}(0)=\delta_{ij}$ we get the 
basic relations between  $\tilde{P}_{ij}(\lambda)$ and 
$\tilde{F}_{ij}(\lambda)$
\begin{equation}
\tilde{P}_{ij}(\lambda)=
\tilde{F}_{ij}(\lambda)\tilde{P}_{jj}(\lambda)+\delta_{ij}
\label{gf2}
\end{equation}
In the following we will call 
$\tilde{P}_i(\lambda)\equiv \tilde{P}_{ii}(\lambda)$ and
$\tilde{F}_i(\lambda)\equiv \tilde{F}_{ii}(\lambda)$.

Before discussing recurrence and transience properties
we briefly recall the definition of the Gaussian model on a graph, whose 
deep relation with random walks will be exploited in the next section.
 
The Gaussian model on the graph $\cal{G}$ can be defined \cite{hhw} 
introducing the field
$\phi_i$ which are the functions of $l^{\infty}(V)=\{(\phi_i)_{i\in V} : 
\sup_i |\phi_{i}|<\infty\}$. It exist a unique Gaussian probability measure 
$d\mu_g{\phi}$ on $l^{\infty}(V)$ with mean zero and covariance 
$(L + \mu)^{-1}$ \cite{hhw} ($\mu_{ij}$ is the diagonal matrix 
$\mu_{ij}=\mu \delta_{ij}$, $\mu>0$); $d\mu_g(\phi)$ characterize the Gaussian 
model and we will write:
\begin{equation}
\langle F(\phi) \rangle =\int F(\phi) d \mu_g(\phi)
\label{gauss}
\end{equation}
and in particular:
\begin{equation}
\langle \phi_i \phi_j \rangle =(L + \mu)^{-1}_{ij}
\label{gauss2}
\end{equation}

Alternately, the Gaussian model can be introduced using standard 
approach of statistical mechanics via the Hamiltonian:
\begin{equation}
{\cal{H}}=\sum_{i,j\in {\cal G}} L_{ij}\phi_i\phi_j + 
\sum_{i\in {\cal G}} \mu \phi_i^2
\label{gauss3}
\end{equation}
together with the Boltzmann weight $\exp(-{\cal H})$, also 
leading to (\ref{gauss2}).

\section{Local recurrence and local transience}

The long time asymptotic behavior of random walks on infinite graphs are 
determined by the large scale topology of the graph and the
quantities $\tilde{F}_i(1)$ and 
$\lim_{\lambda \rightarrow 1}\tilde{P}_i(\lambda)$ can be used to 
characterize the geometry of the graph itself.
In particular a graph is called {\it locally recurrent} if 
\begin{equation}
\tilde{F}_i(1)=1 ~~ {\rm or \  equivalently}~~  
\lim_{\lambda \rightarrow 1} \tilde{P}_i(\lambda)=\infty ~~ \forall i
\label{dric}
\end{equation}
On the other hand if:
\begin{equation}
\tilde{F}_i(1) < 1 ~~ {\rm or \  equivalently}~~  
\lim_{\lambda \rightarrow 1} \tilde{P}_i(\lambda) < \infty
~~ \forall i
\label{dtr}
\end{equation}
the graph is called {\it locally transient}. By standard Markov chains 
properties \cite{woess1} (\ref{dric}) and (\ref{dtr}) are independent from the
site $i$ and then they can be consider as properties of the graphs. 

Let us prove the independence from $i$ of (\ref{dric}).
If $\lim_{\lambda \rightarrow 1} \tilde{P}_i(\lambda)=\infty$ then by 
equation (\ref{gf2}) we get
$\lim_{\lambda\rightarrow 1}\tilde{P}_{ji}(\lambda)=\infty$ for all $j$
($0 <\tilde{F}_{ji}(1)\leq 1$); now from (\ref{pij}) and (\ref{ptij}) we have 
that $z_i P_{ij}(t) =z_j P_{ji}(t)$ and 
$z_i \tilde{P}_{ij}(\lambda) =z_j \tilde{P}_{ji}(\lambda)$. Then also 
$\lim_{\lambda\rightarrow 1}\tilde{P}_{ij}(\lambda)=\infty$ and from 
(\ref{gf2}) we obtain 
$\lim_{\lambda \rightarrow 1}\tilde{P}_{j}(\lambda)=\infty$, $\forall j\in V$.
In an analogous way it can be shown that property (\ref{dtr}) is independent 
from the choice of $i$.

Local transience and local recurrence satisfy  important universality 
properties
\cite{woess1}.
Indeed these properties are not modified if we
substitute the jumping probabilities of the random walker (\ref{pij}) 
with the generalized jumping probability:
\begin{equation}
p_{ij}={J_{ij}\over I_i}~~.
\label{univloc}
\end{equation}
In \cite{woess1} the invariance of the local recurrence properties 
under a wide class of transformations of the graph itself is also proven.
Local recurrence and transience are not modified by 
the addition a finite number of links or the introduction of second neighbor
links on the graph.   Notice that
these basic invariance properties prove that local recurrence and transience 
are determined only by the large scale topology of the graph.

\section{Averages on infinite graphs}

Let us now consider thermodynamic averages on infinite graphs.
The generalized Van Hove sphere $S_{o,r} \subset \cal{G}$ of center $o$ and 
radius $r$ 
is the subgraph of $\cal{G}$ containing all $i \in \cal{G}$ whose chemical 
distance from $o$ $r_{i,o}$ is $\le r$ and all the links of $\cal{G}$ 
joining them. We 
will call $N_{o,r}$ the number of vertices contained in $S_{o,r}$. 

The average in the thermodynamic limit $\bar{\phi}$ of a 
function $\phi_i$ defined on each site $i$ of the infinite graph $\cal{G}$ is:
\begin{equation}
\overline{\phi}\equiv 
\lim_{r\rightarrow\infty} 
{\displaystyle \sum_{i\in S_{o,r}} \phi_i \over \displaystyle N_{o,r}}~~.
\label{deftd}
\end{equation}
The measure $|S|$ of a subset $S$ of 
$V$ is the average value $\overline{\chi(S)}$ of its characteristic 
function $\chi_i(S)$ defined by
$\chi_i(S)=1$ if $i\in S$ and $\chi_i(S)=0$ if $i\not\in S$.
The measure of a subset of links $E'\subseteq E$ is given by:
\begin{equation}
|E'| \equiv 
\lim_{r\rightarrow\infty} 
{\displaystyle E'_r \over  N_{o,r}}~~.
\label{mislinks}
\end{equation}
where $E'_r$ is the number of links of $E'$ contained in the sphere $S_{o,r}$.
The normalized trace $\overline{\rm Tr}B$ of a 
matrix $B_{ij}$ is:
\begin{equation}
\overline{\rm Tr}B \equiv {\overline b}
\end{equation} 
where $b_i \equiv B_{ii}$.
Let us  require that:
\begin{equation}
\lim_{r\to \infty} {|\partial S_{o,r}|\over N_{o,r}}~=~0
\label{isop} 
\end{equation}
where $|\partial S_{o,r}|$ is the number of the vertices of the sphere 
$S_{o,r}$ connected with the rest of the graph. 

Under this hypothesis we now prove that 
the averages of a bounded from below function $\phi_i$ are independent from 
the center $o$ of the spheres sequence, using the fact that 
$\chi_i(S)$ is bounded and that measures of subsets are always well defined. 
From the boundedness of the coordination number we 
get for any couple of vertices $o$ and $o'$:
\begin{equation}
N_{o,r} - (z_{max})^{r_{o,o'}}|\partial S_{o',r}| \leq N_{o',r} 
\leq N_{o,r} + (z_{max})^{r_{o,o'}}|\partial S_{o,r}|
\label{Noo'} 
\end{equation}
and
\begin{equation}
(z_{max})^{-r_{o,o'}}|\partial S_{o,r}| \leq |\partial S_{o',r}|
\leq (z_{max})^{r_{o,o'}}|\partial S_{o,r}|
\label{Noo'b} 
\end{equation}
Let us consider a bounded from below function $\phi_i$. Given 
two vertices $o$ and $o'$, we have:
\begin{equation}
{\displaystyle \sum_{i\in S_{o',r-r_{o,o'}}} \phi_i + 
\sum_{i\in  S_{o,r}\Delta S_{o',r-r_{o,o'}}} \phi_i\over  N_{o,r}} 
= {\displaystyle \sum_{i\in S_{o,r}} \phi_i \over \displaystyle N_{o,r}}=
{\displaystyle \sum_{i\in S_{o',r+r_{o,o'}}} \phi_i - 
\sum_{i\in  S_{o',r+r_{o,o'}} \Delta S_{o,r}} \phi_i\over  N_{o,r}} 
\label{Noo'2}
\end{equation}
where 
$S_{o,r} \subseteq S_{o',r+r_{o,o'}}$, $S_{o',r-r_{o,o'}}\subseteq S_{o,r}$
and $S_{o,r}\Delta S_{o',r-r_{o,o'}}$
is the symmetric difference between $S_{o,r}$ and
$S_{o,r-r_{o,o'}}$. $|S_{o,r}\Delta S_{o',r-r_{o,o'}}|$ denotes
the number of vertices of
$S_{o,r}\Delta S_{o',r-r_{o,o'}}$ and from (\ref{Noo'}) we get:
\begin{equation}
|S_{o,r}\Delta S_{o',r-r_{o,o'}}|\leq
(z_{max})^{r_{o,o'}}|\partial S_{o,r}|
\ \ \ \ \ |S_{o,r}\Delta S_{o',r-r_{o,o'}}|\leq
(z_{max})^{r_{o,o'}}|\partial S_{o,r-r_{o.o'}}|
\label{Noo'3} 
\end{equation}
with an analogous equation holding for $ S_{o',r+r_{o,o'}} \Delta S_{o,r}$.
Defining $\bar{\phi}=0$
if $\phi_i>0$ for all $i$ and $\bar{\phi}=|min_i{\phi}_i|$ otherwise, from 
(\ref{Noo'2}) we have:
\begin{equation}
{\displaystyle \sum_{i\in S_{o',r-r_{o,o'}}} \phi_i - \bar{\phi} 
|S_{o,r}\Delta S_{o',r-r_{o,o'}}| \over  N_{o,r}} 
\leq {\displaystyle \sum_{i\in S_{o,r}} \phi_i \over \displaystyle N_{o,r}}
\leq {\displaystyle \sum_{i\in S_{o',r+r_{o,o'}}} \phi_i + \bar{\phi} 
|S_{o',r+r_{o,o'}} \Delta S_{o,r}| \over  N_{o,r}} 
\label{Noo'4}
\end{equation}
with property (\ref{isop}) of $\cal{G}$ and inequalities (\ref{Noo'3}) we get:
\begin{equation}
\lim_{r\rightarrow\infty}
{\displaystyle \sum_{i\in S_{o',r-r_{o,o'}}} \phi_i \over  N_{o,r}} 
\leq \lim_{r\rightarrow\infty}
{\displaystyle \sum_{i\in S_{o,r}} \phi_i \over \displaystyle N_{o,r}}\leq
\lim_{r\rightarrow\infty}
{\displaystyle \sum_{i\in S_{o',r+r_{o,o'}}} \phi_i \over  N_{o,r}} 
\label{Noo'5}
\end{equation}
since $N_{o,r}=N_{o',r-r_{o,o'}}+|S_{o,r}\Delta S_{o',r-r_{o,o'}}|=
N_{o',r+r_{o,o'}}-|S_{o,r}\Delta S_{o',r-r_{o,o'}}|$ using again (\ref{isop}) 
and (\ref{Noo'3}) we get:
\begin{equation}
\lim_{r\rightarrow\infty}
{\displaystyle \sum_{i\in S_{o',r-r_{o,o'}}} \phi_i \over  N_{o',r-r_{o,o'}}} 
\leq \lim_{r\rightarrow\infty}
{\displaystyle \sum_{i\in S_{o,r}} \phi_i \over \displaystyle N_{o,r}}\leq
\lim_{r\rightarrow\infty}
{\displaystyle \sum_{i\in S_{o',r+r_{o,o'}}} \phi_i \over  N_{o',r+r_{o,o'}}} 
\label{Noo'6}
\end{equation}
Therefore, if the limit with the spheres centered in $o'$ exists, it gives
the same result using as center any vertex $o$.

\section{Recurrence and transience on the average}

The study of thermodynamic properties of statistical models on infinite 
graphs requires the introduction of averages of local quantities. 
The latter are 
related to random walks by the return probabilities on the average 
$\bar{P}$ and $\bar{F}$, which are defined by:
\begin{equation}
\bar{P}=\lim_{\lambda \rightarrow 1}\overline{\tilde{P}(\lambda)}
\label{dbP}
\end{equation}
\begin{equation}
\bar{F}=\lim_{\lambda \rightarrow 1}\overline{\tilde{F}(\lambda)}
\label{dbF}
\end{equation}
A graph $\cal{G}$ is called {\it recurrent on the average} (ROA) 
if $\bar{F}=1$, while it is {\it transient on the average} (TOA) 
when  $\bar{F}<1$.

Recurrence and transience on the average are in general independent of the
corresponding local properties. The first example of this phenomenon 
occurring on inhomogeneous structures was found in a class of infinite
trees called NTD (Fig. 1) which are locally transient but recurrent on the 
average \cite{split}.

Moreover, while for local probabilities (\ref{gf2}) gives:
\begin{equation}
\tilde{P}_i(\lambda)=
\tilde{F}_i(\lambda)\tilde{P}_i(\lambda)+~1
\label{gf22}
\end{equation}
an analogous relation for (\ref{dbF}) and (\ref{dbP}) does not hold since 
averaging (\ref{gf22}) over all sites $i$ would involves the average of
a product, which due to correlations is in general different from the
product of the average. Therefore the double implication 
$\tilde{F}_i(1)=1 \Leftrightarrow \lim_{\lambda \rightarrow 1} 
\tilde{P}_i(\lambda)=\infty$ is not true. Indeed
there are graphs for which $\bar{F}<1$ but $\bar{P}=\infty$ (an example in
shown in Fig. 2) and the 
study of the relation between $\bar{P}$ and $\bar{F}$ is a non trivial 
problem.

\section{Pure and mixed transience on the average}

In this section we study the relation between $\bar{P}$ and $\bar{F}$ and 
we show that a complete picture of the behavior of random walks 
on graphs can be given by dividing transient on the average graphs into 
two further classes, which will be called {\it pure} and {\it mixed} 
transient on the average (TOA).

First, considering a ROA graph, we prove that if $\bar{F}=1$ then 
$\bar{P}=\infty$. In this case for each $\delta >0$ it exists $\epsilon$ such 
that if $1-\epsilon \leq \lambda < 1$, we have: 
$1-\delta \leq \overline{\tilde{F}(\lambda)} \leq 1$. Let us consider the 
subset $S\subseteq V$ of the sites $i$ such that 
$\tilde{F}_i(1-\epsilon)<1-\sqrt{\delta}$ and we call $\bar{S}$ its complement.
We obtain:
\begin{equation}
1-\delta \leq \overline{\tilde{F}(1-\epsilon)} =
\overline{\chi(S)\tilde{F}(1-\epsilon)}
+\overline{\chi(\bar{S})\tilde{F}(1-\epsilon)}\leq (1-\sqrt{\delta})|S|
+|\bar{S}| = 1-\sqrt{\delta} |S|
\label{FP1}
\end{equation}
From (\ref{FP1}) we get $|S|\leq \sqrt{\delta}$ and then 
$|\bar{S}| \geq 1- \sqrt{\delta}$. Exploiting the property that 
$\overline{\tilde{P}(\lambda)}$ is an increasing function of $\lambda$, for 
each $\lambda\geq 1-\epsilon$ we get:
\begin{equation}
\overline{\tilde{P}(\lambda)} \geq \overline{\tilde{P}(1-\epsilon)}\geq
\overline{\chi(\bar{S})(1-\tilde{F}(1-\epsilon))^{-1}}\geq 
|\bar{S}| \delta^{-1/2} \geq (1- \sqrt{\delta})\delta^{-1/2}
\label{FP2}
\end{equation} 
In this way we proved that for arbitrary large value of 
$(1- \sqrt{\delta})\delta^{-1/2}$ ($\delta \rightarrow 0$), it exists 
$\epsilon$ such that for each $\lambda$, $1-\epsilon \leq \lambda < 1$, we have 
$\overline{\tilde{P}(\lambda)}\geq (1- \sqrt{\delta})\delta^{-1/2}$, and 
therefore 
$\bar{P}=\lim_{\lambda\rightarrow 1}\overline{\tilde{P}(\lambda)}=\infty$.

Notice that this proof can be easily generalized to graphs in which there is 
a positive measure subset $S$ such that:
$\lim_{\lambda\rightarrow 1}\overline{\chi(S)\tilde{F}(\lambda)}=|S|$.
Indeed in an analogous way it can be proven that:
\begin{equation}
\bar{P} \geq \lim_{\lambda\rightarrow 1}\overline{\chi(S')\tilde{P}(\lambda)}
=\infty~~~~ \forall S' \subseteq S, |S'|>0 
\label{PFg1}
\end{equation}

We will call {\it mixed} transient on the average a TOA graphs having a 
positive measure subset $S$ such that: 
\begin{equation}
\lim_{\lambda\rightarrow 1}\overline{\chi(S)\tilde{F}(\lambda)}=|S|.
\label{dmixed}
\end{equation}
while a graph will be called {\it pure} TOA, if: 
\begin{equation}
\lim_{\lambda\rightarrow 1}\overline{\chi(S)\tilde{F}(\lambda)} < |S|~~~~
\forall S\subseteq V, |S|>0
\label{dpure}
\end{equation}

Examples of mixed and pure TOA graphs are shown respectively in Fig. 2 and 
Fig. 3. From the previous proof for mixed TOA 
graphs we have $\bar{P}=\infty$; let us now study the behavior of $\bar{P}$ on
pure TOA graphs. We define $k$ as 
\begin{equation}
k = \sup_{S\subseteq V,|S|>0}
\lim_{\lambda\rightarrow 1}\overline{\chi(S)\tilde{F}(\lambda)}|S|^{-1}
\label{FP3}
\end{equation} 
and since the graphs is pure TOA, $k<1$. For each $0<\lambda'< 1$ we 
introduce  $S_{\lambda'}\subseteq V$ as the set of the vertices $i$ such that 
$\tilde{F}_i(\lambda')>k$. Exploiting the property that 
$\tilde{F}_i(\lambda)$ is an increasing function of $\lambda$ we have
$\overline{\chi(S_{\lambda'})\tilde{F}_i(\lambda)}>k|S_{\lambda'}|$ and 
then $\lim_{\lambda\rightarrow 1}
\overline{\chi(S_{\lambda'})\tilde{F}_i(\lambda)}
>k|S_{\lambda'}|$. From (\ref{FP3}) we obtain that $S_{\lambda'}$ has zero 
measure, i. e. it must be $|S_{\lambda'}|=0$. 
Exploiting the definition (\ref{genfun}) we have, 
for all $i\in V$, $\tilde{P}_i(\lambda)\leq (1-\lambda)^{-1}$ and we obtain 
for $\overline{\tilde{P}(\lambda')}$:
\begin{equation}
\overline{\tilde{P}(\lambda')} =
\overline{\chi(\bar{S}_{\lambda'})\tilde{P}(\lambda')}+
\overline{\chi({S}_{\lambda'})\tilde{P}(\lambda')}\leq
\overline{\chi(\bar{S}_{\lambda'})(1-\tilde{F}(\lambda'))^{-1}}+
|S_{\lambda'}|(1-\lambda')^{-1}\leq
|\bar{S}_{\lambda'}|(1-k)^{-1}\leq (1-k)^{-1}
\label{FP4}
\end{equation} 
Taking the limit $\lambda'\rightarrow 1$, we have that 
for pure TOA graphs $\bar{P}$ is finite. 

This prove can be generalized to graphs in which there is a 
positive measure subset $S$ such that for all $S'\subseteq S$, $|S'|>0$, 
$\lim_{\lambda\rightarrow 1}\overline{\chi(S')\tilde{F}(\lambda)}\leq |S'|$
obtaining
\begin{equation}
\lim_{\lambda\rightarrow 1}\overline{\chi(S')\tilde{P}(\lambda)}
< \infty.~~~~ \forall S' \subseteq S, |S'|>0 
\label{PFg2}
\end{equation}

\section{Random walks and infrared properties of the Gaussian model}

The generating function 
$\tilde{P}_i(\lambda)$ is strictly connected with the correlation functions 
of the Gaussian model (\ref{gauss2}) by the following equation:
\begin{equation}
\tilde{P}_i(\lambda)=\sum_{t=0}^{\infty}\lambda^t (Z^{-1}A)^t_{ii}=
(1-\lambda Z^{-1}A)^{-1}_{ii}=
[Z\lambda^{-1}(\Delta + (1-\lambda)\lambda^{-1}Z)^{-1}]_{ii}
\label{ltrg}
\end{equation} 
and taking the limit $\lambda\rightarrow 1$ we have:
\begin{equation}
\lim_{\lambda\rightarrow 1}\tilde{P}_i(\lambda)=
z_i \lim_{\mu\rightarrow 0}(\Delta + \mu Z)^{-1}_{ii}
\label{ltrg2}
\end{equation} 
In \cite{hhw} the invariance of the limit:
$\lim_{\mu\rightarrow 0}(\Delta + \mu )^{-1}_{ii}$ under a local rescaling of 
the masses is proven. In particular we have that if 
$\lim_{\mu\rightarrow 0}(\Delta + \mu )^{-1}_{ii}$ is finite than 
$\lim_{\mu\rightarrow 0}(\Delta + \mu Z)^{-1}_{ii} < \infty$ while when the 
first limit diverges the latter also diverges. Therefore 
on locally recurrent graphs we have 
$\lim_{\mu \rightarrow 0} (\Delta + \mu )^{-1}_{ii} = \infty$ $\forall i$
while for locally transient graphs
$\lim_{\mu \rightarrow 0} (\Delta + \mu )^{-1}_{ii} < \infty$ $\forall i$.

Let us now consider the average generating function. From (\ref{ltrg}) we get:
\begin{equation}
\lim_{\lambda\rightarrow 1}\overline{\tilde{P}(\lambda)}=
\lim_{\mu\rightarrow 0}\overline{\rm Tr}[Z(\Delta + \mu Z)^{-1}]
\label{ltrg3}
\end{equation} 
and since the connectivity of $\cal{G}$ is bounded we get:
\begin{equation}
\lim_{\mu\rightarrow 0}\overline{\rm Tr}(\Delta + \mu Z)^{-1} \leq
\lim_{\lambda\rightarrow 1}\overline{\tilde{P}(\lambda)}\leq
z_{max} \lim_{\mu\rightarrow 0}\overline{\rm Tr}(\Delta + \mu Z)^{-1}
\label{ltrg4}
\end{equation} 
Exploiting the universality properties of the Gaussian model \cite{univ}, we 
have that $\lim_{\mu\rightarrow 0}\overline{\rm Tr}(\Delta + \mu Z)^{-1}$ is 
finite if and only if 
$\lim_{\mu\rightarrow 0}\overline{\rm Tr}(L + \mu )^{-1}<\infty$, where $L$ is 
a generalized Laplacian given by (\ref{defL}). Finally from inequalities
(\ref{ltrg4}) we get that 
$\lim_{\mu\rightarrow 0}\overline{\rm Tr}(L + \mu )^{-1}=
\lim_{\mu\rightarrow 0}\overline{\langle \phi_i \phi_i \rangle}$ diverges
if $\bar{P}=\infty$ i.e. on ROA and mixed TOA graphs, while it is finite 
on pure TOA graphs, where $\bar{P}<\infty$.

\section{Separability and statistical independence}

In this section we prove and discuss an important property characterizing 
mixed TOA graphs which allows to simplify the study of statistical
models on these very inhomogeneous structures. 
We will show that in this case the graph $\cal{G}$ can be always decomposed 
in a pure TOA subgraph $\cal{S}$ and a ROA subgraph $\bar{\cal{S}}$ with
independent jumping probabilities by 
cutting a zero measure set of links $\partial {\cal{S}} \equiv \{ (i,j)\in E |
i \in {\cal{S}} \and j \in \bar{\cal{S}} \}$.
The separability property implies that the two subgraphs
are statistically independent and that their thermodynamic  properties
can be studied separately. Indeed, the partition functions referring 
to the two subgraphs factorize \cite{bcv}.    

As a first step, from definition (\ref{dmixed}) the set of vertices $V$ of 
a mixed TOA graph $\cal{G}$ can always be decomposed in two 
complementary subsets $S$ and $\bar{S}$ such that
\begin{equation}
{\overline{\chi(S')\tilde{F}(1)} \over |S'|}<1
\label{cl1}
\end{equation}
 for all $S'\subseteq S$ with $|S'|>0$ and
\begin{equation}  
{\overline{\chi({S}'')\tilde{F}(1)} \over |{S}''|}=1
\label{cl2}
\end{equation}
for all ${S}''\subseteq \bar{S}$ with $|{S}''|>0$.

To this decomposition we can associate the two subgraphs $\cal{S}$ and 
$\bar{\cal{S}}$ defined as follows: $\cal{S}$ has $S$ as set of vertices and
its links are all the links $(i,j) \in \cal{G}$ such that $i,j\in S$; in the
same way $\bar{\cal{S}}$ has $\bar S$ as set of vertices and its links 
are all the links $(i,j) \in \cal{G}$ such that $i,j\in {\bar S}$.
Let us now  prove that the measure of the boundary $|\partial \cal{S}|$
(\ref{mislinks}) is zero.

We introduce $B_S$, the border set of $\cal{S}$, defined as the set of the 
vertices $i\in S$ with $(i,j)\in \partial {\cal{S}}$ for some $j$ 
while we will call $B_{\bar{S}}$ the border set of $\bar{\cal{S}}$.
Proving that $|\partial {\cal{S}}|=0$ is equivalent to show that the measure 
of $B_S$ and $B_{\bar{S}}$ is zero. Indeed we have 
$|\partial {\cal{S}}|_r \leq |B_S|_r \leq z_{max} |\partial {\cal{S}}|_r$
and 
$|\partial {\cal{S}}|_r \leq |B_{\bar{S}}|_r \leq 
z_{max} |\partial {\cal{S}}|_r$,
where $|B_S|_r$ and $|B_{\bar{S}}|_r$ are the number of sites in $B_S$ and 
$B_{\bar{S}}$ contained in the sphere $S_{o,r}$.

Let us suppose that $\partial {\cal{S}}\geq 0$ and that $|B_S|\geq0$, 
$|B_{\bar{S}}|\geq 0$.
From (\ref{PFg1}) and (\ref{PFg2}) we have: 
\begin{equation}
\lim _{\lambda\rightarrow 1}
\overline{\chi(B_S) \tilde{P}(\lambda)}\leq \infty
\label{ppure}
\end{equation}
and 
\begin{equation}
\lim _{\lambda\rightarrow 1}
\overline{\chi(B_{\bar{S}}) \tilde{P}(\lambda)}= \infty
\label{proa}
\end{equation}

We will now derive a relation between 
$\overline{\chi(B_S) \tilde{P}(\lambda)}$ and
$\overline{\chi(B_{\bar{S}}) \tilde{P}(\lambda)}$ which can not be satisfied 
if (\ref{ppure}) and (\ref{proa}) hold, leading to a contradiction.
This implies that $|\partial {\cal{S}}|=0$.

Let us evaluate $\tilde{P}_i(\lambda)$ in a site $i\in B_S$
\begin{equation}
\tilde{P}_i(\lambda)  =  \sum_t \lambda^t p_{ii}^t 
 =  \sum_t \lambda^t \sum_{jk}p_{ik} p_{kj}^{t-2}p_{ji}
 \geq  \sum_t \lambda^t \sum_{j\in B_{\bar{S}}}p_{ij} p_{jj}^{t-2}p_{ji}
\label{cl3}
\end{equation}
where in the inequality we do not consider  the terms in which $j\not =k$ and 
$j \not \in B_{\bar{S}}$
Exploiting the fact that $p_{ij}\geq 1/ z_{max}$
we get:
\begin{equation}
\tilde{P}_i(\lambda) 
\geq  {\lambda^{2} \over z_{max}^2} 
\sum_t \lambda^{t-2} \sum_{j\in B_{\bar{S},i}}  p_{jj}^{t-2}
= {\lambda^{2} \over z_{max}^2}  \sum_{j\in B_{\bar{S},i}} 
\tilde{P}_{jj}(\lambda)
\label{cl4}
\end{equation}
where $B_{\bar{S},i}$ is the set of the nearest neighbors sites of $i$ which 
belong to $B_{\bar{S}}$. If we take the average over the sites $i\in B_S$ 
we have:
\begin{equation}
\overline{\chi(B_S)\tilde{P}(\lambda)} 
\geq  {\lambda^{2} \over z_{max}^2}  \lim_{r \rightarrow \infty} 
{ \chi_i(B_S) \over N_{o,r}} 
\sum_{i\in S_{o,r}}\sum_{j\in B_{\bar{S},i}}  \tilde{P}_{jj}(\lambda)
\geq  {\lambda^{2} \over z_{max}^2}  \lim_{r \rightarrow \infty} 
{ \chi_j(B_{\bar{S}}) \over N_{o,r}} 
\sum_{j\in S_{o,r}}  \tilde{P}_{jj}(\lambda)= {\lambda^{2} \over z_{max}^2}
\overline{\chi(B_{\bar{S}})\tilde{P}(\lambda)} 
\label{cl5}
\end{equation}
If we take the limit $\lambda\rightarrow 1$ we have:
\begin{equation}
\lim_{\lambda \rightarrow 1}
\overline{\chi(B_S)\tilde{P}(\lambda)} \geq  {1 \over z_{max}^2}
\lim_{\lambda \rightarrow 1}\overline{\chi(B_{\bar{S}})\tilde{P}(\lambda)} 
\label{cl6}
\end{equation}
Expressions (\ref{cl6}), (\ref{ppure}) and (\ref{proa})
can not be satisfied at the same time and therefore one must have
$|\partial {\cal{S}}|=0$

Finally we have to prove that $\cal{S}$ is a pure TOA graph and $\bar{\cal{S}}$
is a ROA graph, i.e. we introduce the restricted jumping probability on $S$ 
and $\bar{S}$ $p_{ij}^{S}$ and $p_{ij}^{\bar{S}}$, given by  
$p_{ij}^{S}=p_{ij}$ if $i,j \in S$, $p_{ij}^{S}=0$ otherwise and an analogous 
definition for $p_{ij}^{\bar{S}}$. Then we show that $S$ and 
$\bar{S}$ with the new jumping probabilities $p_{ij}^{S}$ and 
$p_{ij}^{\bar{S}}$ are respectively pure TOA and ROA.

More generally for a 
walker on $\cal{S}$ starting from $i$, we call $P^{S}_{ij}(t)$ the probability
 of reaching site $j$ in $t$ steps and  
$F^{S}_{ij}(t)$ the probability of reaching $j$ for the first time in $t$ 
steps. We will prove that: 
\begin{equation}
\overline{\tilde{P}^{S}(\lambda)}=
\overline{\sum_{t=0}^{\infty}\lambda^t P^{S}_{i}(t)}=
\overline{\chi(S)\tilde{P}(\lambda)}|S|^{-1}~~~~~
\overline{\tilde{F}^{S}(\lambda)}=
\overline{\sum_{t=0}^{\infty}\lambda^t F^{S}_{i}(t)}=
\overline{\chi(S)\tilde{F}(\lambda)}|S|^{-1}
\label{cl7}
\end{equation}
where the average of $\tilde{P}^{S}(\lambda)$ and of $\tilde{F}^{S}(\lambda)$ 
is taken considering $\cal{S}$ as the whole graph. Analogous 
equations hold also for $\bar{\cal{S}}$. From (\ref{cl7}), (\ref{cl1}) and 
(\ref{PFg2}) we easily obtain
that $\bar{P}^{S} < \infty$ i.e. $\cal{S}$ is pure TOA, while if we call 
$P^{\bar{S}}_{ij}(t)$ and $F^{\bar{S}}_{ij}(t)$ the probabilities for a 
random walk on
$\bar{\cal{S}}$, we get $\bar{F}^{\bar{S}}=1$, i.e. $\bar{\cal{S}}$ is ROA.

To prove equations (\ref{cl7}) first we have to show that:
\begin{equation}
\overline{\tilde{P}^{S}(\lambda)}=
\sum_{t=0}^{\infty}\lambda^{t} \overline{P^{S}(t)}~~~~~ \lambda<1
\label{cl8}
\end{equation}
Equation (\ref{cl8}) implies that the thermodynamic average and the sum over 
the discretized times $t$ commute when $\lambda<1$.
To prove (\ref{cl8}) notice that for all $\lambda < 1$ we have:
\begin{equation}
\overline{\tilde{P}^{S}(\lambda)}=
\lim_{r\rightarrow\infty} \sum_{i\in S_{o,r}} N_{o,r}^{-1}\left(
\sum_{t=0}^{\bar{t}}\lambda^{t} P^{S}_{i}(t) +
\sum_{t=\bar{t}}^{\infty}\lambda^{\bar t} P^{S}_{i}(t)\right)=
\sum_{t=0}^{\bar{t}}\lambda^{t} \overline{P^{S}(t)} + 
\lim_{r\rightarrow\infty} \sum_{i\in S_{o,r}} N_{o,r}^{-1}
\sum_{t=\bar{t}}^{\infty}\lambda^{ t} P^{S}_{i}(t)
\label{cl9}
\end{equation} 
Now $\sum_{i\in S_{o,r}} N_{o,r}^{-1} 
\sum_{t=\bar{t}}^{\infty}\lambda^{\bar t} P^{S}_{i}(t)
\leq \lambda^{\bar{t}}(1-\lambda)^{-1}$ and letting in (\ref{cl9}) 
$\bar{t}\rightarrow \infty$ we get (\ref{cl8}). Obviously an analogous 
equation holds also for $ F^{S}_{i}(t)$, $P_{i}(t)$ and $F_{i}(t)$. Then 
we can prove (\ref{cl7}) showing that:
\begin{equation}
\overline{{P}^{S}(t)}=\overline{\chi(S){P}(t)}|S|^{-1}~~~~~
\overline{{F}^{S}(t)}=\overline{\chi(S)\tilde{F}(t)}|S|^{-1}
\label{cl10}
\end{equation}
We define  $d(i,B_S)$ as the 
distance between $i$ and the cutset $B_S$: 
$d(i,B_S)=\inf_{k\in B_S} r_{i,k}$ and will call $S_t$ the subset of $S$ 
such that: $S_t=\{ i\in S | d(i,B_S) \leq t \}$, exploiting the boundedness 
of the coordination number we get:
\begin{equation}
|S_t|<(z_{max})^t |B_S|=0
\label{cl11}
\end{equation}  
since $ |B_S|=0$.
Taking the average of $P^{S}_{i}(t)$ we have:
\begin{equation}
\overline{{P}^{S}(t)}
=\overline{\chi(\bar{S}_t)P^{S}(t)}+\overline{\chi({S}_t)P^{S}(t)}
\label{cl12}
\end{equation}
Now $\overline{\chi({S}_t)P^{S}(t)}\leq |S_t|=0$, and then 
$\overline{{P}^{S}(t)}
=\overline{\chi(\bar{S}_t)P^{S}(t)}$. Finally exploiting the fact that on 
$\bar{S}_t$ we have  $P^{S}_{i}(t)= P_{ii}(t)$, we obtain (\ref{cl10}).
Following analogous steps we obtain the equality for 
$\overline{{F}^{S}(t)}$ and for the averages $\overline{{P}^{\bar{S}}(t)}$
and $\overline{{F}^{\bar{S}}(t)}$ defined on $\bar{S}$.

\section{Discussion and conclusions}

In this paper we have presented a systematic mathematical analysis of the Type Problem
for random walks on infinite graphs by considering 
return probabilities averaged over all sites. After showing that {\it recurrence and 
transience on the average} (ROA and TOA) do not in general coincide with the corresponding local 
properties, we prove that TOA has 
to be splitted in two complementary subcases, the {\it pure} and the {\it mixed} one.
Then we show that a mixed TOA graph can always be decomposed in a ROA and a pure TOA subgraphs
by cutting a zero measure set of links.
This property has deep physical implications, since it 
allows to decompose a statistical model defined on a mixed TOA graph in two thermodynamically
independent models defined respectively on the ROA and pure TOA subgraphs. 

In conclusion, we introduced an exhaustive classification of infinite networks in terms
on their average recurrence and transience properties, stating the Type Problem on the Average.
This classification is the relevant one in the study of thermodynamic properties of
statistical models on inhomogeneous structures.

%%%%%%%%%%%%%%%%%%%%%%%%%%%%%%%%%%%%%%%%%

\newpage

\begin{center}
\epsfig{file=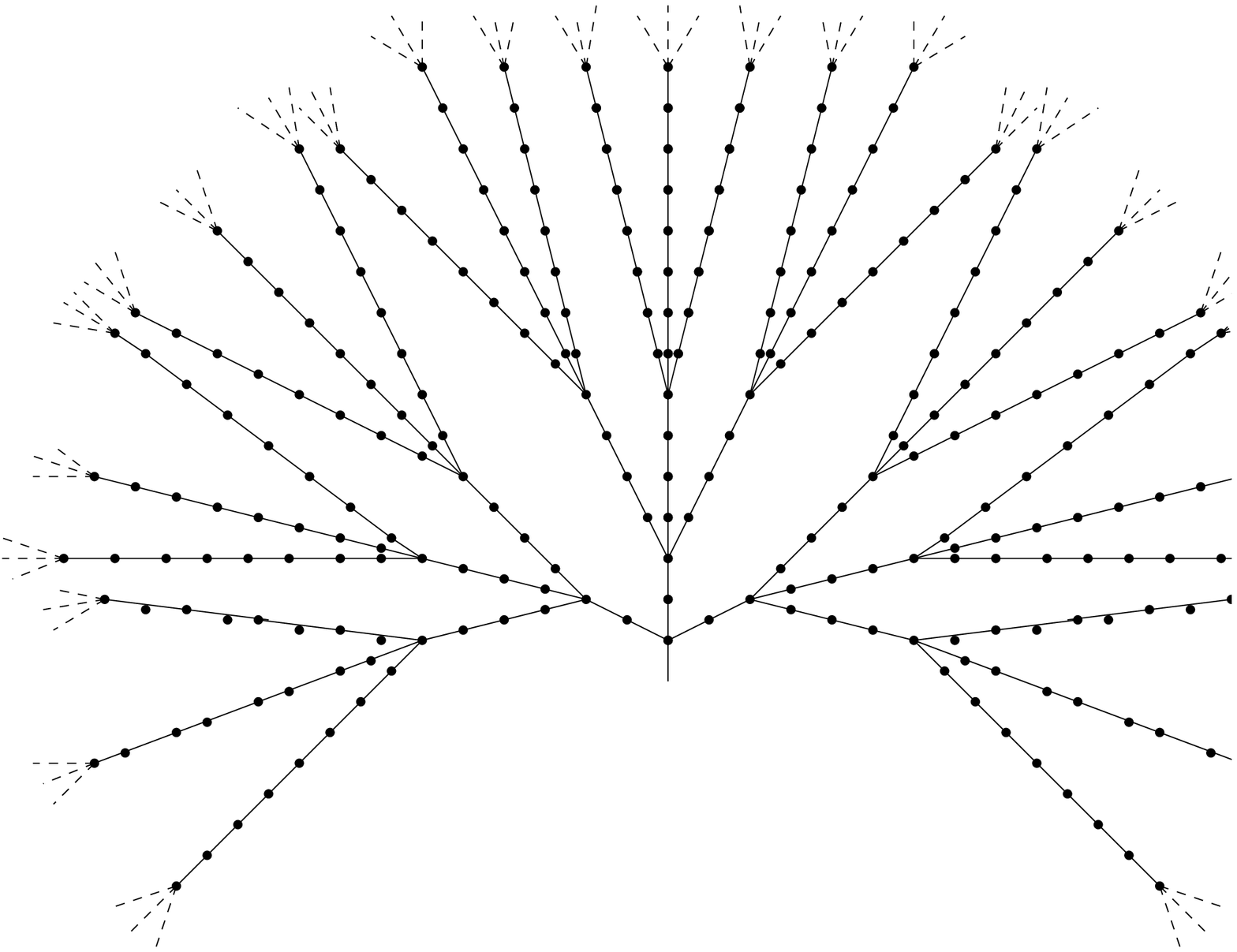,height=6cm}
\end{center}
{Fig. 1: The NTD tree: the distances between the ramifications increase 
exponentially. This graph is locally transient and recurrent on the average.}

\begin{center}
\epsfig{file=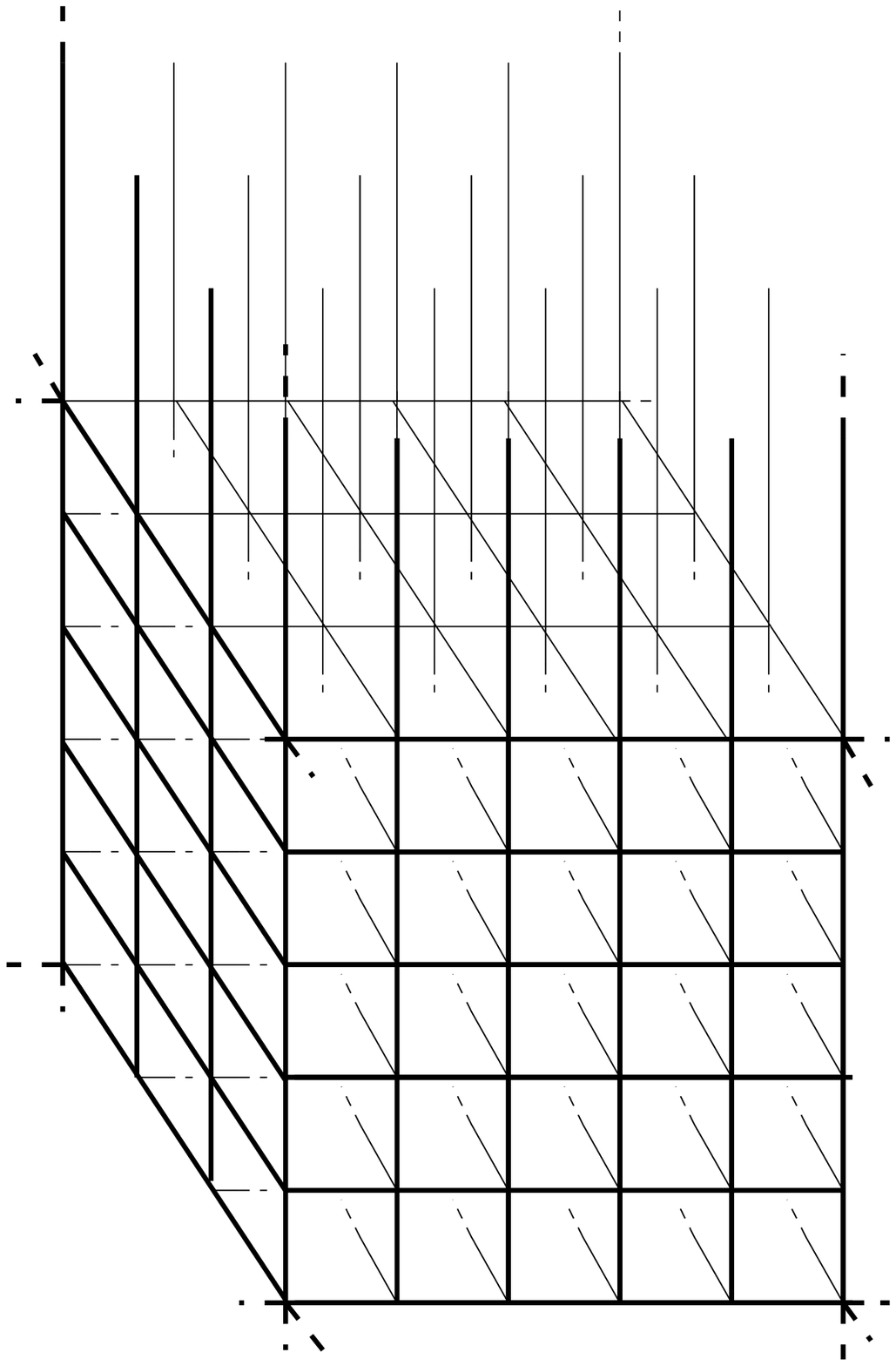,height=6cm}
\end{center}
{Fig  2: A mixed TOA graph: the cubic lattice is a pure TOA graph while the 
hairs are ROA.}

\begin{center}
\epsfig{file=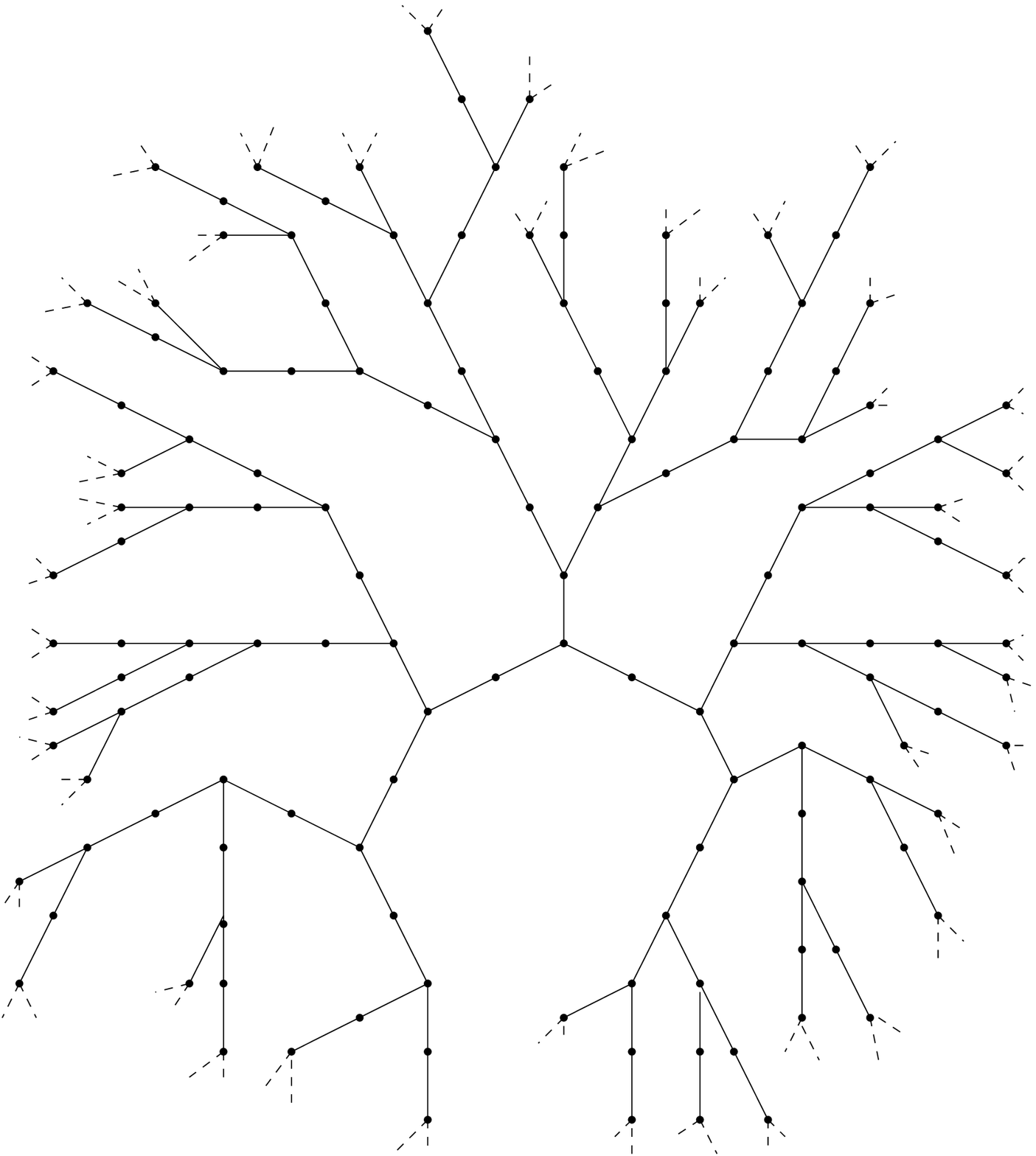,height=6cm}
\end{center}
{Fig  3: A pure TOA graph, i.e. an inhomogeneous Bethe  lattice in which the 
distance between ramifications can be 1 or 2.}

\end{document}